\documentclass[conference]{IEEEtran}
\ifCLASSINFOpdf

\else

\fi

\usepackage{graphicx}
\usepackage{amsthm}
\usepackage{subfigure}
\usepackage{amssymb, amsmath, amsfonts, epsfig, latexsym, times}
\usepackage{bm}
\usepackage{cite}
\usepackage{flushend}
\usepackage{booktabs}
\usepackage{multirow}

\usepackage{algorithmic}
\usepackage{algorithm}

\begin{document}
%
\bibliographystyle{IEEEtran}

\title{\huge{Heterogeneous Services Provisioning in Small Cell Networks with Cache and Mobile Edge Computing}}

\author{\IEEEauthorblockN{Zhiyuan Tan\IEEEauthorrefmark{1}, F.~Richard~Yu\IEEEauthorrefmark{2}, Xi Li\IEEEauthorrefmark{1}, Hong Ji\IEEEauthorrefmark{1},
and Victor C.M. Leung\IEEEauthorrefmark{3}}
\IEEEauthorblockA{\IEEEauthorrefmark{1}Key Lab. of Universal Wireless Commun., Ministry of Education, Beijing Univ. of Posts and Telecomm., Beijing, China}
\IEEEauthorblockA{\IEEEauthorrefmark{2}Depart. of Systems and Computer Eng., Carleton Univ., Ottawa, ON, Canada}
\IEEEauthorblockA{\IEEEauthorrefmark{3}Depart. of Electrical and Computer Eng., the Univ. of British Columbia, BC, Canada}
}

\maketitle

\begin{abstract}
In the area of full duplex (FD)-enabled small cell networks, limited works have been done on consideration of cache and mobile edge communication (MEC). In this paper, a virtual FD-enabled small cell network with cache and MEC is investigated for two heterogeneous services, high-data-rate service and computation-sensitive service. In our proposed scheme, content caching and FD communication are closely combined to offer high-data-rate services without the cost of backhaul resource. Computing offloading is conducted to guarantee the delay requirement of users. Then we formulate a virtual resource allocation problem, in which user association, power control, caching and computing offloading policies and resource allocation are jointly considered. Since the original problem is a mixed combinatorial problem, necessary variables relaxation and reformulation are conducted to transfer the original problem to a convex problem. Furthermore, alternating direction method of multipliers (ADMM) algorithm is adopted to obtain the optimal solution. Finally, extensive simulations are conducted with different system configurations to verify the effectiveness of the proposed scheme.
\end{abstract}


\IEEEpeerreviewmaketitle

\section{Introduction}
In recent years, with the development of smart phones, more and more heterogeneous services and related applications are emerging, such as high-speed multi-media, interactive gaming and AR/VR applications. The high-data-rate service and computation-sensitive service become two typical heterogeneous services. And the requirements with high quality of experience (QoE) make the shortage of spectrum resource become especially prominent. However, with recent advances in self-interference (SI) cancellation technologies, full duplex (FD) technology becomes an effective solution to improve the utilization of spectrum \cite{SSG14,LYJ15}, which has attracted a lot of attentions and has been applied to other communication technologies, such as relay, small cell networks (SCNs) and device-to-device (D2D) communications\cite{ZZC15,CYL13,CYJ16,ZYLW15}. Although the system performance is significantly improved, how to satisfy different QoE requirements of heterogeneous services is very challenging. Only limited works have been done to combine FD technology with caching and mobile edge computing (MEC).

The rapid development of caching technology brings a new thought to high-data-rate services. Applying caching to heterogeneous network nodes can change the way of passive and reactive content request and transmission in the traditional communication network. Popular contents can be proactively predicted and cached in the storage facilities, which shortens the distance between users and contents and offers high-QoE services \cite{IW16}. Besides, MEC is regarded as another promising technology to computation-sensitive services, which enables cloud-computing capabilities in radio access networks (RANs) in proximity to mobile subscribers.
With MEC, the challenge about the relationship between resource-hungry applications and resource-constrained mobile devices can be overcome \cite{XLLF16,HYH16}. More importantly, the QoE, especially delay, can be obviously improved when the computing tasks are offloading to MEC.

Although caching and MEC can bring a lot of benefits, how to efficiently integrate and manage different system resources becomes a challenge.
Network function virtualization (NFV) is an effective approach to manage physical network infrastructures and wireless resources \cite{LY15,LYZ15,LY15m}. Through NFV, system resources can be dynamically and flexibly shared by users with differentiated services improving the resource utilization.

So, in this paper, we investigate the virtualized FD-enabled small cell networks framework with cache and MEC, where two heterogeneous services are considered.
The joint resource allocation problem is formulated. The distinct features of this paper are summarized as follows.
\begin{itemize}
  \item {We design a novel virtualized FD-enabled small cell networks framework with caching and MEC. From the perspective of heterogeneous services, we aim at high-data-rate and computation-sensitive services. And NFV is enabled to guarantee the feasibility and flexibility of the framework. Moreover, FD and caching can complement to offer high-data-rate service saving backhaul resource.  }
  \item {Different from existing works, in this paper, we formulate a joint virtual resource allocation problem, where user association, power control, caching and computing offloading policies and joint resource allocation are taken into account, which satisfies users' QoE requirements.}
  \item {Due to the non-convexity of original problem, we transfer the original problem to a convex problem by variables relaxation and reformulation. A distributed algorithm is adopted to obtain the optimal solution with low complexity. Simulation results show the superiority of our scheme.}
\end{itemize}


\section{Network Framework and System Model}
\subsection{Network Framework}
As illustrated in Fig. 1 (a), the virtualized FD-enabled small cell networks framework consists of three layers.
In the mobile network virtualized operator (MVNO) layer, there are a lot of MVNOs which can lease different physical infrastructures and resources through NFV. Access layer consists of one macro base station (MBS) and some SBSs with different cache and MEC. Corresponding, two kinds of users with different service requirements are deployed in the user layer. In this framework, users do not have the authorization to access the SBSs directly. Instead, they receive the virtualized services from the MVNOs. For simplicity, user mobility \cite{YL01} and handover \cite{MYL04,MYL07,YK07} are not considered in this paper. Each MVNO is corresponding to a SBS with cache and computing hardware. According to users' requests, with the management of MVNOs, the SBSs and related resources can finally be allocated to users.

As shown in Fig. 1 (b), there are two kind of content delivery ways for users with high-data-rate service (named Service I). The SBSs can directly communicate with users when required contents are cached, like User 1. Otherwise, users will request the content to the MBS by FD communication, like User 2.
Thus, caching and FD communication can efficiently supplement each other to serve the users with Service I. As for users with computation-sensitive service (named Service II), computing tasks can be offloaded to the MEC to execute with low delay when with good channel conditions and reasonable resource allocation. 
\begin{figure}[tp]
\centering
{\includegraphics[width=0.5\textwidth]{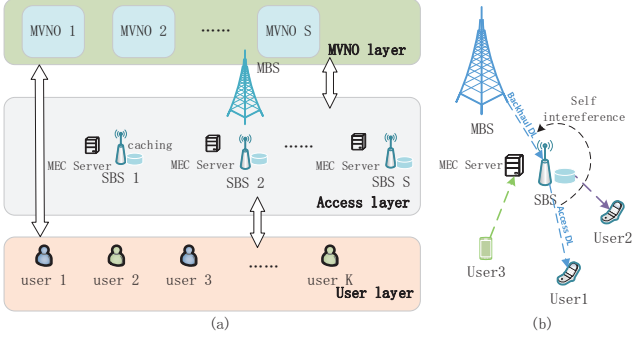}}
\caption{A virtualized FD-enabled small cell networks with cache and MEC.}
\label{fig.sys}
\end{figure}
\subsection{System Model}
In our framework, let $\mathcal{S}=\{1,...,S\}$ denote the set of SBSs. According to two main service requirements, the $K$ users can be classified into two classes.  And the corresponding user sets are denoted $\Omega_A=\{1,...K_1\}$ and $\Omega_B=\{K_1+1,...,K\}$, respectively.
In this paper, the frequency division duplex (FDD) is adopted, where the downlink (uplink) spectrum is allocated for users with Service I (II). The corresponding bandwidths are denoted as $B_d$ and $B_u$. Meanwhile, the orthogonal spectrum scheme is adopted which enables no interference among users in the DL and UL transmissions.
We first consider the content direct delivery, the achievable spectrum efficiency of UE $i$ in SBS $k$ in the downlink is shown as
\begin{equation}
\label{eq: achievable spectrum efficiency of UE in downlink}
\ r_{ik}=\log_2 (1+\frac{p_{ki}h_{ki}}{\sigma^2}),
\end{equation}
where the $p_{ki}$ is the transmission power of SBS $k$ for user $i$ and the $h_{ki}$ is the channel gain from the SBS $k$ to user $i$.

When contents are not cached, FD communication is adopted. The SBS receives data from the MBS while it transmits data to its users at the same time in the same frequency band with SI.
Then for the access downlink in FD communication, the spectrum efficiency of user $i$ in SBS $k$ is same as the one in direct delivery, which can be also denoted as $r_{ik}$. In the backhaul downlink, the spectrum efficiency of user $i$ is expressed as
\begin{equation}
\label{eq: achievable spectrum efficiency of UE in backhaul downlink}
\ r_{ik}^b=\log_2 (1+\frac{a_{ik}P_mh_{ik}^m}{\mathrm{SI}+\sigma^2}),
\end{equation}
where $P_m$ and $a_{ik}$ are total power of MBS and the fraction of power allocated to the link with SBS $k$ and UE $i$, respectively. $h_{ik}^m$ is the corresponding channel gain. $\mathrm{SI}=\varpi p_{ki}$ is residual SI where $\varpi$ is the residual SI gain \cite{CYJL16}.

Caching model includes the current content caching and content refreshment. Let $hit_{ik}\in\{0,1\}$ be the current content caching indicator. And $hit_{ik}=1$ means that the content requested by UE $i$ has been cached in SBS $k$, $hit_{ik}=0$ means opposite. We assume that the popularity of contents (request probability of contents) follows Zipf distribution \cite{LCXS2016}.
The popularity of the $f$th popular content out of the overall $F$ types contents can be described as $p_f=\frac{Q}{f^{\mu}}$, where the parameter $\mu\geq1$. And $ Q=1/\sum\limits_{f=1}^{F}\frac{1}{f^\mu}.$
In addition, we introduce binary variable $c_{ik}\in\{0,1\}$ as the content refreshment indicator. Namely, $c_{ik}=1$ indicates that the SBS $k$ caches the content requested by user $i$ and $c_{ik}=0$ otherwise. Please note that $c_{ik}=0$ if $hit_{ik}=1$ to avoid repeated caching.

Then a computing offloading model is proposed for users with Service II. We assume each computing task can not be further divided and can be described as $M_j=\{R_j,D_j\}$ where $R_j$ is the size of input data to MEC and $D_j$ denotes the computing capability required for this task which is quantized by the number of CPU cycles \cite{ZML16}. Due to the fact that computing tasks can be executed locally or be offloaded to the MEC server, the relevant overheads are discussed.

Local Execution: Let $w_j^l$ denote the local computing capability (CPU cycles). The total execution time for UE $j$ can be given as $t_j^l=D_j/w_j^l$.

Computing Offloading: Let $w_k^c$ denote the MEC computing capability in SBS $k$. And $z_{jk}\in[0,1]$ is the fraction indicator of computing resource of SBS $k$ allocated to UE $j$. Thus, the execution time $t_{jk}^e$ in MEC for UE $j$ is $t_{jk}^e=D_j/z_{jk}w_k^c$. And uplink transmission time is $t_{jk}^u=R_{j}/y_{jk}B_k^ur_{jk}$ where denominator item is transmission rate of user $j$. The total time to complete the task $M_j$ in MEC is denoted as $t_{j}^c=t_{jk}^u+t_{jk}^e$.

\section{Problem Formulation}
\subsection{Constraints}
Let $x_{ik}\in\{0,1\}$ denote the association indicator for users with Service I, where $x_{ik}=1$ indicates that user $i$ chooses to access to the SBS $k$, otherwise, $x_{ik}=0$. Similarly, for users with Service II, $x_{jk}\in\{0,1\}$ denotes the computing offloading indicator (association variable).
Then user association constraints are given as follows.
\begin{equation}
\label{eq: association choice of UE with Service I}
\ C1: \sum \limits_{k\in\mathcal{S}}x_{ik}\leq 1, \forall i\in\Omega_A.
\end{equation}
\begin{equation}
\label{eq: association choice of UE with Service II}
\ C2: \sum \limits_{k\in\mathcal{S}}x_{jk}\leq 1, \forall j\in\Omega_B,
\end{equation}
Constraints 1 and 2 show every user with Service I/II can only be served by a SBS.

Let $y_{ik} \in [0,1]$ and $y_{jk}\in[0,1]$ denote the fraction indicator of the frequency bandwidth allocated to user $i$ or user $j$ associated with SBS $k$ in the downlink and uplink transmission. Then the downlink and uplink bandwidth constraints can be formulated as
\begin{equation}
\label{eq: bandwidth constraints of UE with Service I}
\ C3: \sum \limits_{k\in\mathcal{S}} \sum \limits_{i\in\Omega_A}x_{ik}y_{ik}\leq 1.
\end{equation}
\begin{equation}
\label{eq: bandwidth constraints of UE with Service II}
\ C4: \sum \limits_{k\in\mathcal{S}} \sum \limits_{j\in\Omega_B}x_{jk}y_{jk}\leq 1.
\end{equation}

As for users with Service I, the requirement for high data rate $R_i^d$ should be satisfied. For transmissions locally or in the access downlink, the transmission rate constraint is formulated as
\begin{equation}
\label{eq: data rate constraint of UE with Service I}
\ C5: \sum \limits_{k\in\mathcal{S}} x_{ik}y_{ik}B_dr_{ik}\geq R_i^d, \forall i\in \Omega_A.
\end{equation}

In addition, FD communication requires that transmission rate in the backhaul downlink is higher than the one in the access downlink, namely $r_{ik}^b \geq r_{ik}$ when the required content is not cached. So we can get
\begin{equation}
\label{eq: data rate constraint in FD commmunication of UE with Service I}
\ C6: x_{ik}a_{ik} \geq \frac{x_{ik}(1-hit_{ik})p_{ki}h_{ki}(\varpi p_{ki}+\sigma^2)}{P_mh_{ik}^m \sigma^2} , \forall k, i,
\end{equation}
where FD communication is adopted when $hit_{ik}=0$.

In FD communication, the MBS power is limited so that only limited backhaul links are served.  And $a_{ik}\in [0,1]$ is the fraction indicator of the power of MBS. So we can get
\begin{equation}
\label{eq: power constraint of UE with Service I}
\ C7: \sum\limits_{k\in \mathcal{S}}\sum \limits_{i\in \Omega_A} a_{ik} \leq 1.
\end{equation}

As for Service II, the constraint that delay in MEC should be less than the minimal computing delay $t_0$ ($t_0\leq t_j^l$), that is $t_{j}^c\leq t_0, \forall j\in \Omega_B$. Since $t_j^c$ is the sum of transmission delay and execution delay, we extend the delay requirement to two constraints to avoid one of them in $t_{j}^c$ to occupy too much resulting to unfair and unreasonable resource allocation.
The transmission delay threshold is $T_u$, and the execution delay is less than the difference value $t_0-T_u$.  The detailed constraints are formulated as C8 and C9.
\begin{equation}
\label{eq: transmission delay 1 constraint of UE with Service II}
\ C8: t_{jk}^u\leq T_u, \forall j\in \Omega_B.
\end{equation}

\begin{equation}
\label{eq: transmission delay 2 constraint of UE with Service II}
\ C9: t_{jk}^e\leq t_0-T_u, \forall j\in \Omega_B.
\end{equation}

Another constraint is the size of all cached contents in every SBS in the content refreshment period should not exceed the storage capability. Then the caching storage constraint is formulated as
\begin{equation}
\label{eq: caching storage constraint of UE with Service I}
\ C10: \sum \limits_{i\in\Omega_A} x_{ik}c_{ik}d_{ik}\leq D_k, \forall k\in \mathcal{S}.
\end{equation}
where the $D_k$ is the left storage capability of SBS $k$. And the $d_{ik}$ is the size of content requested by user $i$ associated with SBS $k$. And $c_{ik}\in \{0,1\}$ is the cache refreshment indicator.

Similarly, each MEC server has a limited computing capability and $z_{ik}\in [0,1]$ is the fraction indicator of the computing resource. Then the computing constraint is formulated as
\begin{equation}
\label{eq: computing capability constraint of UE with Service II}
\ C11: \sum \limits_{j\in\Omega_B} x_{jk}z_{jk}\leq 1, \forall k\in \mathcal{S}.
\end{equation}

\subsection{Utility Function}
Considering that the different performance criteria in two heterogeneous services lead to different revenues and costs, we use the subtraction between system revenues and costs as the system utility function.
In Service I, the saved backhaul band contains two parts: part one from local caching or FD communication and part two from caching refreshment policy, which is denoted as
\begin{equation}
\label{eq: computing ability constraint of UE with Service II}
\ C_{ik}=x_{ik}y_{ik}B_dr_{ik}+x_{ik}c_{ik}(1-hit_{ik})v_{ik}.
\end{equation}
where $v_{ik}=q_{ik}d_{ik}$ is the average size of the cached content with refreshment policy.

In addition, we can get the consumed band, cache and power for user $i$ which are denoted as $G_{ik}=x_{ik}y_{ik}B_d$, $Cache_{ik}=x_{ik}c_{ik}(1-hit_{ik})s_{ik}$ and $Q_{ik}=(1-hit_{ik})a_{ik}^mP_m$, respectively. And $s_{ik}$ is the size of required content.
In order to maximize the total utility of all MVNOs, it is necessary to efficiently allocate the virtual resources and guarantee the fairness during the resource allocation process. Then we apply the logarithmic function to the utility function as the method in \cite{CYJL16} since the logarithmic function is a nondecreasing and concave function. $R_{ik}$ and $R_{jk}$ are the sum rate for user $i$ and $j$. Then we define $R_{ik}'=\log R_{ik}$, $C_{ik}'=\log C_{ik}$, $R_{jk}'=\log R_{jk}$. Thus, the utility function for user $i$ can be formulated as
\begin{align}
\label{eq: Utility function of UE I}
u_{ik}&=\alpha_iR_{ik}'+\beta_iC_{ik}'-\gamma_iG_{ik}-\eta_iCache_{ik}-\varepsilon_iQ_{ik}\nonumber \\
&=\alpha_ix_{ik}\log y_{ik}B_dr_{ik}-\gamma_ix_{ik}y_{ik}B_d \nonumber \\
&+\beta_ix_{ik}\log (y_{ik}B_dr_{ik}+c_{ik}(1-hit_{ik})v_{ik}) \nonumber \\
&-\eta_ix_{ik}c_{ik}(1-hit_{ik})s_{ik}-\varepsilon_i(1-hit_{ik})a_{ik}^mP_m
\end{align}
where $\alpha_i$ and $\beta_i$ denote revenue coefficient per unit of received data rate and saved backhaul resource. We define that $\gamma_{i}$, $\varepsilon_i$, $\eta_i$
are the price coefficient of downlink communications for consuming system bandwidth, MBS power in the FD self-backhaul transmission and caching, respectively.

Similarly, for Service II, the revenues are from transmission rate of input data and the saving energy consumption of users at the cost of the frequency bandwidth and computing resource. With the increasingly rigid environmental standards and rising energy costs, there are great interests on the energy issues in wireless networks \cite{XYJL12,BYC12,YZX11,BYY15}. Therefore, we need to model energy consumption. We denote $\psi_j$ and $\theta_j$ as the the revenue coefficient per unit of transmitting the computing data and saving energy for user $j$. And $\vartheta_j$ and $\phi_j$ are the price coefficient per unit of consumed computing resource and frequency band. Thus, the utility function of user $j$ with Service II is given as
\begin{align}
\label{eq: Utility function of UE II}
u_{jk}=&\psi_j x_{jk}\log y_{jk}B_jr_{jk}+\theta_j x_{jk}\frac{D_j}{w_j^l}p_j^l-\phi_j x_{jk}y_{jk}B_u \nonumber\\
&-\vartheta_j x_{jk}z_{jk}w_k^c
\end{align}
where $p_j^l$ is the power of user $j$ to execute the computing task locally.
In consequence, the resource allocation problem is formulated as
\begin{equation}
\begin{aligned}
\label{formulated problem}
&~~~~~~\max\limits_{(\mathbf{x_i},\mathbf{x_j},\mathbf{y_i},\mathbf{y_j},\mathbf{c},\mathbf{z},\mathbf{a})} \sum \limits_{k\in \mathcal{S}}(\sum \limits_{i\in\Omega_A}u_{ik}+\sum\limits_{j\in\Omega_B}u_{jk}). \\
&~~~~~~~~~\textmd{s.t.} ~~C1-C11.\\
\end{aligned}
\end{equation}
\section{Problem Reformulation and Solution}
\subsection{ Problem Reformulation}
Firstly, the variable relaxation is conducted. We relax the binary variables $x_{ik}$, $x_{jk}$ and $c_{ik}$ to the continuous variables, that is $x_{ik}\in[0,1]$, $x_{jk}\in[0,1]$ and $c_{ik}\in[0,1]$. The relaxed variables can be regarded as the sharing in the time scale.

We define $\tilde{y}_{ik}=x_{ik}y_{ik}$, $\tilde{c}_{ik}=x_{ik}c_{ik}$, $\tilde{a}_{ik}=x_{ik}a_{ik}$, $\tilde{y}_{jk}=x_{jk}y_{jk}$, and $\tilde{z}_{jk}=x_{jk}z_{jk}$.
Taking $\tilde{y}_{ik}=x_{ik}y_{ik}$ for example, we can see $y_{ik}=\frac{\tilde{y}_{ik}}{x_{ik}}$ except $x_{ik}=0$ due to the loss of definition at $x_{ik}=0$. To guarantee one-to-one mapping in the variable substitution process, we complement the definition when $x_{ik}=0$ where definition is presented as
\begin{equation}
y_{ik}=\begin{cases}
 \frac{\tilde{y}_{ik}}{x_{ik}}& \textrm{if $x_{ik}>0$},\\
0& \textrm{if $otherwise$}.
\end{cases}
\end{equation}
Then in the same way, we can complement the definition of $y_{jk}$, $c_{ik}$, $a_{ik}$, and $z_{jk}$ to conform to one-to-one mapping of variables. Then the original problem is reformulated as (19).

Based on variable substitution above, the convexity of optimization problem (19) can be proved. Firstly, in terms of the objective function, some similar structures like $x_{ik}\log\frac{\tilde{y}_{ik}}{x_{ik}}$ are well-known perspective function of logarithmic function \cite{boyd2009convex}. Based on related works \cite{CYJL16}, the convexity can be obtained.
The convexity of the perspective function keeps consistent with the original function and the function $\log \tilde{y}_{ik}$ is concave about $\tilde{y}_{ik}$, so $x_{ik}\log{\tilde{y}_{ik}/x_{ik}}$ is concave. Similarly,  $x_{ik}\log{\tilde{c}_{ik}/x_{ik}}$, $x_{ik}\log{\tilde{a}_{ik}/x_{ik}}$, $x_{jk}\log{\tilde{y}_{jk}/x_{jk}}$, $x_{jk}\log{\tilde{z}_{jk}/x_{jk}}$ are concave.
With the theory that a negative concave problem is a convex problem, we can conclude that the form of the objective function in (19) is a linear sum of convex problems. In consequence, the optimization problem (19) is a convex problem.
\begin{equation}
\begin{aligned}
&~~~~~~\max\limits_{(\mathbf{x_i},\mathbf{x_j},\mathbf{\tilde{y}_i},\mathbf{\tilde{y}_j},\mathbf{\tilde{c}},\mathbf{\tilde{z}},\mathbf{\tilde{a}})} \sum \limits_{k\in \mathcal{S}}(\sum \limits_{i\in\Omega_A}u_{ik}+\sum\limits_{j\in\Omega_B}u_{jk}). \\
&~~~~~~~~~\textmd{s.t.} ~~C1,C2.\\
&~~~~~~~~~C3':\sum \limits_{k\in\mathcal{S}} \sum \limits_{i\in\Omega_A}\tilde{y}_{ik}\leq 1.\\
&~~~~~~~~~C4':\sum \limits_{k\in\mathcal{S}} \sum \limits_{j\in\Omega_B}\tilde{y}_{jk}\leq 1.  \\
&~~~~~~~~~C5':\sum \limits_{k\in\mathcal{S}}\tilde{y}_{ik}B_k^dr_{ik}\geq R_i^d, \forall i\in \Omega_A.\\
&~~~~~~~~~C6': \tilde{a}_{ik} \geq \frac{x_{ik}(1-hit_{ik})p_{ki}h_{ki}(\varpi p_{ki}+\sigma^2)}{P_mh_{ik}^m \sigma^2} , \forall i,k.\\
&~~~~~~~~~C7': \sum\limits_{k\in \mathcal{S}}\sum \limits_{i\in \Omega_A} (1-hit_{ik})\tilde{a}_{ik} \leq 1.\\
&~~~~~~~~~C8':r_{jk}^u=\tilde{y}_{jk}B_k^ur_{jk}\geq \frac{R_j}{T_u},\forall j\in \Omega_B.\\
&~~~~~~~~~C9':\frac{D_j}{x_{jk}w_{jk}}=\frac{D_j}{\tilde{z}_{jk}w_k}\leq{t_0-T_u},\forall j\in \Omega_B.\\
&~~~~~~~~~C10':\sum \limits_{i\in\Omega_A} \tilde{c}_{ik}d_{ik}\leq D_k, \forall k\in \mathcal{S}.  \\
&~~~~~~~~~C11':\sum \limits_{j\in\Omega_B} \tilde{z}_{jk}\leq 1, \forall k\in \mathcal{S}.
\end{aligned}
\end{equation}

\subsection{Problem Solution}
Due to the constraints, optimization variables are not separable with respect to each SBS. In order to achieve a distributed optimization algorithm, necessary decoupling measures are conducted. Therefore, we introduce $\mathbf{\hat{x}_i}^k$, $\mathbf{\hat{x}_j}^k$, $\mathbf{\hat{y}_i}^k$, $\mathbf{\hat{y}_j}^k$ and $\mathbf{\hat{a}}^k$ as the local copies of global association indicator for SBS $k$, which can be interpreted as the SBSs' opinion about the corresponding global variable.
Besides, the consistence between all local variables and the associated global variables should be held. Namely, $\hat{x}_{in}^k=x_{in}$, $\hat{y}_{in}^k=\tilde{y}_{in}$, $\hat{a}_{in}^{mk}=\tilde{a}_{in}^m$, $\forall i,k,n$ and $\hat{x}_{jn}^k=x_{jn}$, $\hat{y}_{jn}^k=\tilde{y}_{jn}$, $\forall j,k,n$. By introducing the local vectors $\mathbf{\hat{x}_i}^k$, $\mathbf{\hat{x}_j}^k$, $\mathbf{\hat{a}}^k$, $\mathbf{\hat{y}_i}^k$ and $\mathbf{\hat{x}_j}^k$ into (19), the feasible set of local vectors for each SBS $k$ can be defined as $\Phi_k$.

Then we can get the local utility function for SBS $k$ as
\begin{equation}
g_k=\begin{cases}
-(\sum \limits_{i\in\Omega_A}u_{ik}'+\sum\limits_{j\in\Omega_B}u_{jk}')  & \textrm{$\mathbf{\hat{x}_i}^n,\mathbf{\hat{y}_i}^n,\mathbf{\tilde{c}},\mathbf{\hat{x}_j}^n,\mathbf{\hat{y}_j}^n,
\mathbf{\tilde{z}},\mathbf{\hat{a}} \in \Phi_k $ }\\
0& \textrm{$otherwise$}
\end{cases}
\end{equation}

Therefore, we can get the global consensus problem of problem (19) as follows
\begin{equation}
\begin{aligned}
&\min G(\mathbf{\hat{x}_i}^k,\mathbf{\hat{y}_i}^k,\mathbf{\tilde{c}},\mathbf{\hat{x}_j}^k,\mathbf{\hat{y}_j}^k,
\mathbf{\tilde{z}},\mathbf{\hat{a}})=\sum \limits_{k\in \mathcal{S}} g_k(\mathbf{\hat{x}_i}^k,\mathbf{\hat{y}_i}^k,\mathbf{\tilde{c}},\mathbf{\hat{x}_j}^k,\mathbf{\hat{y}_j}^k,
\mathbf{\tilde{z}},\mathbf{\hat{a}})\\
&~~~~~~~~~\textmd{s.t.} ~\hat{x}_{in}^k=x_{in}, \hat{y}_{in}^k=\tilde{y}_{in}, \hat{a}_{in}^k=a_{in}, \forall i,k,n.\\
&~~~~~~~~~~~~~\hat{x}_{jn}^k=x_{jn}, \hat{y}_{jn}^k=\tilde{y}_{jn}, \forall j,k,n.
\end{aligned}
\end{equation}

It can be seen that the objective function is separable across SBSs. However, the global variables involved in the consensus constraints couple the problem with respect to the SBSs.  Alternating direction method of multipliers (ADMM) algorithm is applied to solve this problem in a distributed way. ADMM is an effective distributed algorithm to solve the convex problem with quick convergence, which is suitable to the features in our problem. First, an augmented Lagrangian is formulated with respect to the consensus constraints. The augmented Lagrangian is formulated as (22).

\begin{figure*}[!t]
\normalsize
\begin{align}
&\mathfrak{L}_\rho(\{\mathbf{\hat{x}_i}^k,\mathbf{\hat{y}_i}^k,\mathbf{\hat{x}_j}^k,\mathbf{\hat{y}_j}^k,\mathbf{\tilde{c}},
\mathbf{\tilde{z}},\mathbf{\hat{a}}^k\}_{k\in \mathcal{S}},\{(\mathbf{x_i}, \mathbf{\tilde{y}_i}, \mathbf{a}, \mathbf{x_j}, \mathbf{\tilde{y}_j}\},\{\bm{\lambda}_k, \bm{\nu}_k, \bm{\xi}_k, \bm{\varsigma}_k, \bm{\omega}_k\}) = \sum \limits_{k\in\mathcal{S}}g_k(\mathbf{\hat{x}_i}^k,\mathbf{\hat{y}_i}^k,\mathbf{\hat{x}_j}^k,\mathbf{\hat{y}_j}^k,\mathbf{\tilde{c}},
\mathbf{\tilde{z}},\mathbf{\hat{a}}^k) \nonumber\\
&+\sum\limits_{i\in \Omega_A}\sum\limits_{n,k\in \mathcal{S}}\lambda_{in}^k(x_{in}^k-x_{in})+\frac{\rho}{2}\sum\limits_{i\in \Omega_A}\sum\limits_{n,k\in \mathcal{S}}(x_{in}^k-x_{in})^2+\sum\limits_{i\in \Omega_A}\sum\limits_{n,k\in \mathcal{S}}\nu_{in}^k(y_{in}^k-y_{in})+\frac{\rho}{2}\sum\limits_{i\in \Omega_A}\sum\limits_{n,k\in \mathcal{S}}(y_{in}^k-y_{in})^2 \nonumber\\
&+\sum\limits_{i\in \Omega_A}\sum\limits_{n,k\in \mathcal{S}}\xi_{in}^k(a_{in}^{k}-a_{in})+\frac{\rho}{2}\sum\limits_{i\in \Omega_A}\sum\limits_{n,k\in \mathcal{S}}(a_{in}^{k}-a_{in})^2 + \sum\limits_{j\in \Omega_B}\sum\limits_{n,k\in \mathcal{S}}\varsigma_{in}^{k}(x_{jn}^k-x_{jn})+\frac{\rho}{2}\sum\limits_{j\in \Omega_B}\sum\limits_{n,k\in \mathcal{S}}(x_{jn}^k-x_{jn})^2 \nonumber\\
&+\sum\limits_{i\in \Omega_B}\sum\limits_{n,k\in \mathcal{S}}\omega_{in}^k(y_{jn}^k-y_{jn})+\frac{\rho}{2}\sum\limits_{i\in \Omega_B}\sum\limits_{n,k\in \mathcal{S}}(y_{jn}^k-y_{jn})^2.
\end{align}
\hrulefill
\end{figure*}
And $\mathbf{\lambda}_k, \mathbf{\nu}_k, \mathbf{\xi}_k, \mathbf{\varsigma}_k, \mathbf{\omega}_k$ are the vector of Lagrange multipliers and penalty parameter $\rho \in \mathbb{R}_{++}$ is a positive constant parameter to determine the convergence speed. 
The ADMM method is composed of sequential optimization phases over the primal variables followed by the method of multipliers update for the dual variables. The system optimal solutions are calculated by sequential iteration for local variables, global variables and dual variables. The detailed steps are as follows: By applying the ADMM to problem (21), we first minimize the augmented Lagrangian in (22) over the local variables, accordingly, perform global variables update, and finally, the dual-variable is updated. After several iterations, ADMM will converge to the optimal solution.

\section{ Simulation Results and Discussions}
In this section, the performance in our proposed scheme is demonstrated by comparing with the followed schemes: (a) Scheme 1 only with (w.) FD but without (w.o.) caching; (b) Scheme 2 only with caching but without FD. Meanwhile, the performance is considered from the number of users, the number of SBSs and hit ratio. And the metric of performance is measured through the system utility and system cost.

In the simulation, we consider a single cell covered with $1000m*1000m$. One MBS is located at the center of the cell, SBSs and users with a random deployment. The transmit power of MBS, SBSs, and users are $46dBm$, $20dBm$ and $10dBm$, respectively. And the available bandwidth of downlink and uplink are both $20MHz$. For simplicity, assume that the numbers of users with Service I and II are same (named the number of user pair). And caching is cheaper than the cost of power in FD communication.
Then we take $\alpha_i=4*10^6 units/bps$, $\beta_i=5*10^6 units/bps$,  $\gamma_i=10units/Hz$, $\varepsilon_i=10^7units/W$, $\eta_i=2*10^6units/Mb$, $\forall i\in \Omega_A$ and $\psi_j=2*10^6 units/bps$, $\theta_j=2*10^8 units/Joule/s$, $\phi_j=10 units/Hz$ and  $\vartheta_j=2.7*10^6units/GHz$, $\forall j\in \Omega_B$.

Fig. 2 reveals the impact of the number of user pairs on system utility in the three different schemes. As illustrated in Fig. 2, system utility increases with the number of user pairs increasing. But the trend becomes slow. At first, the system resources are enough to allocate to users, with the increasing of users, resources including the number of SBSs, frequency band, caching, computing, and MBS power become the constraint to system utility. In addition, we can see our proposed scheme is superior to the other two schemes. The gaps of utility value among different curves mainly come from Service I where FD and caching can guarantee users served without backhaul band. By contrast, Scheme 1 only achieve FD communication by using more expensive power resource. And Scheme 2 inevitably depend on the bakhaul resource due to different caching ratio. Moreover, the performance of Scheme 1 is superior to Scheme 2. This is because that fewer contents are cached when the number of SBSs is small.
\begin{figure}[!t]
\centering
{\includegraphics[width=0.47\textwidth]{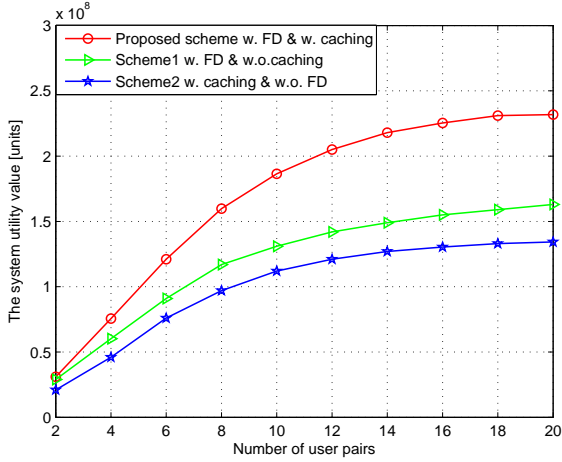}}
\caption{System utility versus the number of user pairs (the number of SBSs=4, hit ratio=0.3).}
\end{figure}
\begin{figure}[!t]
\centering
{\includegraphics[width=0.47\textwidth]{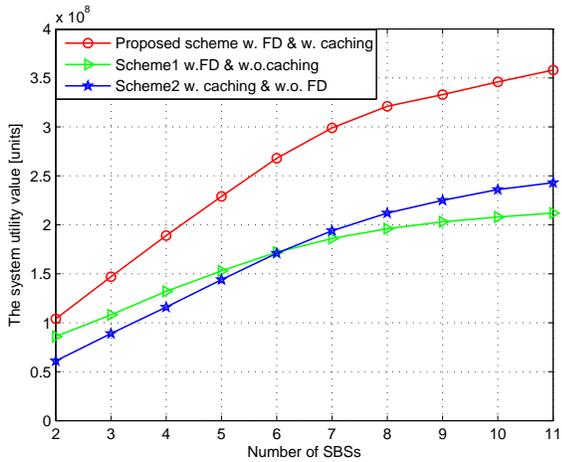}}
\caption{System utility versus the number of SBSs (the number of user pairs=10, hit ratio=0.3).}
\end{figure}

The effect of number of SBSs on system performance is revealed in Fig. 3. The value of system utility increases as the number of SBSs increases and our proposed scheme is superior to the other two schemes. This is because more dense deployment of SBSs can efficiently improve user association and resource allocation, and more users can be better served in the downlink and uplink. And the trend of increasing becomes slow since the number of users is limited. When the number of users is large enough, system resource becomes the constraint to system utility. Besides, we can find that Scheme 1 is superior to Scheme 2 at first. With the number of SBSs increasing, Scheme 2 is better than Scheme 1. It is because that more contents can be cached to serve users locally with the number of SBSs increasing.

%

\section{Conclusions and Future Work}
In this paper, we presented a novel virtual small cell network framework. Then we proposed a joint scheme with respect to two main heterogeneous services. FD communication and cache were integrated closely to effectively save the backhaul resource. Then we formulated the user association, power control, caching and computing policies and joint resource allocation problem. Necessary variables relaxation and reformulation were conducted to guarantee the convexity of optimize problem. A distributed algorithm was applied to obtain the optimal solution with low complexity. Simulations results showed that our proposed scheme is superior to the other two schemes from the aspects of system utility and system cost. Future work is in progress to consider non orthogonal multiple access in the proposed scheme.

\section*{Acknowledgment}
This work is jointly supported by National Natural Science Foundation of China (Grant Nos. 61671088 and 61501047).

\ifCLASSOPTIONcaptionsoff
  \newpage
\fi

\bibliography{Ref}

\end{document}